\documentclass[11pt,a4paper]{article}
\usepackage[utf8]{inputenc}
\usepackage[T1]{fontenc}
\usepackage{lmodern}
\usepackage[margin=1in]{geometry}
\usepackage{graphicx}
\usepackage{booktabs}
\usepackage{array}
\usepackage{enumitem}
\usepackage[hidelinks,breaklinks=true]{hyperref}
\usepackage{url}
\usepackage{caption}
\begin{document}

\title{\bfseries Post-Deployment Accountability in AI Governance: A Cross-Regulatory Empirical Analysis of AI Incidents}
\author{Ummara Mumtaz  $\cdot$  Summaya Mumtaz}
\date{}
\maketitle

\begin{abstract}
Post-deployment accountability has become central to AI governance, yet little empirical evidence shows whether monitoring, incident reporting, and impact assessment obligations are visible when AI systems fail. This study analyzes real-world AI incidents from the AI Incident Database (2020--2026) and codes them against nine post-deployment provisions from the EU AI Act, the NIST AI Risk Management Framework, and the GDPR. The findings show substantial accountability gaps: 77.1\% of incidents lack evidence of EU AI Act post-market monitoring, and 99.6\% lack documented Data-Protection Impact Assessment evidence. Governance gaps are also systemic, with 9.8\% of incidents simultaneously non-compliant under two or more regimes. Incidents detected through internal monitoring show much higher compliance than externally detected incidents (87.5\% vs 5.3\% under the EU AI Act; 95.8\% vs 58.1\% under NIST), suggesting that monitoring capacity is a key condition for effective post-deployment governance. Building on these findings, the paper proposes the Proactive AI Governance Compliance Framework (PAGCF), a four-phase lifecycle for pre-deployment assessment, continuous monitoring, incident preparedness, and cross-framework verification.

\end{abstract}


\section*{1. Introduction}
Artificial-intelligence (AI) systems are now embedded in high-stakes decisions across healthcare, criminal justice, finance, transportation, hiring, and public administration [20, 2]. Their governance problem is therefore no longer limited to forecast model design or abstract ethical alignment. It also concerns the institutional capacity to observe deployed systems, detect harms, preserve evidence, notify affected actors, and alter or withdraw systems when failures occur. In response, regulators have converged on a specific accountability strategy: post-deployment governance. The EU AI Act [5] requires post-market monitoring and serious-incident reporting for high-risk systems; the NIST AI Risk Management Framework [18] calls for continuous monitoring, risk response, communication, and controlled decommissioning; and the General Data Protection Regulation [4] imposes safeguards against solely automated decisions, mandates breach notification, and requires data-protection impact assessments for high-risk processing. Together, these instruments define the dominant blueprint through which AI harms are expected to be detected, documented, reported, and remediated after deployment [23, 24, 9].

Whether this blueprint works in practice is a different question. Much AI governance scholarship remains principally normative: it clarifies which principles should apply [7, 6], how risk should be classified [19, 10], or how accountability should be structured [8, 27, 3]. A smaller empirical literature examines audits, practitioner constraints, and incident databases, but it rarely evaluates whether specific regulatory obligations are triggered when deployed AI systems cause harm. Prior AIID studies [11, 26] have characterised the ethical dimensions of incidents, and recent work has piloted automated governance assessment [17]; however, existing research does not systematically map individual incidents to provision-level obligations across multiple governance regimes. This paper addresses that gap by treating AI incidents as observable stress tests of post-deployment accountability.

Contributions. We make four contributions:

\begin{enumerate}[leftmargin=1.4em,itemsep=1pt,topsep=2pt]
\item A provision-level, cross-regulatory empirical analysis of 480 real-world AI incidents (2020--2026), coded against nine post-deployment provisions of the EU AI Act, NIST AI RMF, and GDPR.
\item Provision-level evidence that governance failure is systemic: the same incidents tend to escape multiple frameworks simultaneously.
\item Identification of internal monitoring as a strong compliance correlate (17$\times$ higher EU AI Act compliance in internally detected incidents), with explicit discussion of selection-bias threats to a causal interpretation.
\item The Proactive AI Governance Compliance Framework (PAGCF), a four-phase, provision-mapped lifecycle methodology accompanied by a risk-stratified tiering scheme and an implementation checklist.
\end{enumerate}

The study is guided by four research questions:

\begin{enumerate}[leftmargin=1.4em,itemsep=1pt,topsep=2pt]
\item RQ1. To what extent do real-world AI incidents trigger the accountability mechanisms prescribed by the EU AI Act, NIST AI RMF, and GDPR?
\item RQ2. How do governance gaps vary across frameworks, sectors, risk categories, and geographies?
\item RQ3. What incident characteristics distinguish cases in which governance mechanisms function from those in which they fail?
\item RQ4. How can observed failure patterns inform a proactive governance methodology aimed at reducing accountability gaps before incidents occur?
\end{enumerate}

The remainder of the paper is organised as follows. Section 2 reviews the AI-governance literature and the three regulatory regimes examined. Section 3 describes the sampling strategy and coding scheme. Sections 4 and 5 present the empirical findings, including the cross-regulatory gap analysis and the internal-monitoring effect. Section 6 introduces the PAGCF. Sections 7 and 8 discuss implications and threats to validity, and Section 9 concludes.

\section*{2. Related Work and Regulatory Background}
Ethics-based AI governance has proliferated rapidly. Jobin, Ienca, and Vayena [7] identify 84 AI ethics guidelines worldwide with broad convergence around transparency, fairness, non-maleficence, responsibility, and privacy, while Hagendorff [6] documents a substantial gap between principle and practice. Mittelstadt et al. [14] anticipated many of the ethical challenges now regulated, and Selbst et al. [22] formalise the difficulty of extracting fairness from its socio-technical context. A parallel line of work examines the mechanics of algorithmic accountability. Kroll et al. [8] argue that legal accountability for algorithmic outcomes need not depend on model transparency, Wieringa [27] systematises the sociotechnical prerequisites of accountable AI, and Cobbe, Veale, and Singh [3] extend the analysis to algorithmic supply chains.

Operational tools have emerged in parallel: internal audits [20], model cards [13], ethics-based auditing [15], and algorithmic impact assessments [12]. Empirical studies of practitioners [21, 1] report that even mature organisations struggle to translate governance intent into operational routineechoing the concept of governance theatre in which policy artefacts multiply without corresponding enforcement. This body of work establishes two premises for the present study. First, accountability is not reducible to explainability or documentation; it depends on organisational routines that make harms discoverable and actionable. Second, those routines are difficult to observe directly. Incident reports therefore provide an imperfect but valuable empirical window into whether governance infrastructures leave detectable traces when systems fail. The AI Incident Database (AIID) [11] provides a systematic repository of real-world AI failures and has become the primary empirical substrate for governance research. Wei and Zhou [26] use the AIID to characterise ethics-related incidents; Nach [17] pilots LLM-based automated governance assessment. Neither maps incidents to specific regulatory provisions across multiple regimes at the provision level, which is the approach adopted here.

\subsection*{2.1 Regulatory background}
EU AI Act. Regulation (EU) 2024/1689 [5] is the first comprehensive legal framework specifically designed to regulate AI systems. Adopted in June 2024, it imposes tiered obligations based on risk and is central to post-deployment governance through Articles 72--73. Article 72 requires providers of high-risk AI systems to implement post-market monitoring, including ongoing surveillance, performance-data collection, and updated technical documentation. Article 73 mandates the reporting of serious incidents involving death, serious health harm, serious property damage, or fundamental-rights violations within specified timeframes. Legal commentators [24, 9] note that these provisions represent the sharpest post-deployment obligations in any current AI-specific regime.

NIST AI Risk Management Framework. NIST AI RMF 1.0 [18] is a voluntary, process-oriented framework organised around four functions: GOVERN, MAP, MEASURE, and MANAGE. For post-deployment governance the MANAGE function is most relevantMANAGE 1 (ongoing monitoring), MANAGE 2 (risk response), MANAGE 3 (communication and transparency), and MANAGE 4 (decommissioning). We also include GOVERN 4 on organisational accountability. Unlike the EU AI Act's prescriptive legal model, the NIST RMF relies on voluntary organisational processes, raising empirical questions about whether this approach yields comparable governance outcomes.

General Data Protection Regulation. Although the GDPR [4] is not AI-specific, several provisions are central to post-deployment AI governance. Article 22 protects individuals from solely automated decision-making with legal or similarly significant effects and requires safeguards including human oversight, explanation, and contestation [25]. Articles 33--34 require breach notification to supervisory authorities within 72 hours and, where applicable, to affected data subjects. Article 35 requires Data-Protection Impact Assessments (DPIAs) for high-risk processing. Because the GDPR combines extraterritorial reach with an established enforcement regimeincluding fines of up to 4\% of global annual turnoverit offers a valuable comparative case for assessing whether stronger enforcement capacity yields more effective post-deployment governance.

\subsection*{2.2 The governance gap}
Despite the proliferation of AI governance frameworks, it remains unclear whether post-deployment accountability mechanisms function effectively when AI systems cause harm. The missing evidence is not simply another catalogue of AI risks, but an account of whether the obligations designed to manage those risks are activated in practice. Existing research lacks three elements: (i) provision-level cross-framework comparison using large-scale incident data, (ii) quantitative evidence on whether internal monitoring is associated with improved compliance outcomes, and (iii) a proactive governance methodology grounded in observed failure patterns rather than in purely normative design. This study addresses each of these elements while preserving a cautious distinction between documented compliance, absence of evidence, and legally adjudicated non-compliance.

\section*{3. Methodology}
\subsection*{Data source and sampling}
Our analysis draws on the AI Incident Database (AIID), a publicly accessible repository of real-world AI failures and harms curated by the Responsible AI Collaborative [11]. The full snapshot contained 1,365 incident records. We restrict the analysis to incidents between 2020 and 2026, a period in which post-deployment AI governance moved from soft-law principles toward concrete legal and organisational obligations. This temporal boundary also reduces the risk of comparing incidents across substantially different regulatory environments: earlier incidents often predate the policy vocabulary now used to describe monitoring, incident reporting, and impact assessment. The sample was refined in three stages:

\begin{enumerate}[leftmargin=1.4em,itemsep=1pt,topsep=2pt]
\item Regex-assisted screening + manual review, retaining incidents relevant to post-deployment AI governance (n = 1,123).
\item Conceptual screening, excluding malicious-misuse and pure-misinformation cases (n = 568).
\item Two rounds of manual review removing scams, impersonation, copyright/IP disputes, and other cases with limited conceptual relevance, yielding the final analytic sample of n = 480 incidents.
\end{enumerate}

Each record includes structured metadata (deployment sector, technology type, severity indicators, geographic location, date of occurrence, and detection method), associated report texts, and links to complementary taxonomies from CSET, the MIT AI Risk Repository, and the Global Monitoring Framework (GMF). The resulting sample should therefore be understood as a sample of publicly visible AI governance failures, not as a complete census of all deployed-system harms. That distinction is central to the interpretation of the results below.

\subsection*{3.2 Coding scheme}
We operationalise compliance indicators for nine post-deployment provisions across the three frameworks using automated content analysis over incident report texts and structured metadata. The objective is not to substitute for formal legal adjudication, but to identify whether incident narratives contain evidence that the governance mechanisms contemplated by each framework were present, absent, or not publicly documented. Although less nuanced than expert legal assessment, this approach enables systematic provision-level coding at a scale that would be impractical to complete manually and is comparable to the automated governance assessment strategy used in Nach [17]. For each provision an incident is assigned to one of four categories: compliant, partially compliant, non-compliant, or insufficient evidence. Compliance is assessed only for the subset of incidents to which a provision is applicable.

\begin{table}[t]
\centering
\footnotesize
\caption{Operationalised regulatory provisions and coding approach.}
\begin{tabular}{@{}p{0.18\textwidth} p{0.24\textwidth} p{0.27\textwidth} p{0.23\textwidth}@{}}
\toprule
Framework & Provision & Compliance Indicator & Coding Approach \\
\midrule
EU AI Act & Art. 72 (Post-Market Monitoring) & Evidence of ongoing monitoring & Detection method + monitoring keywords \\
EU AI Act & Art. 73 (Serious Incident) & Formal incident reporting & Severity flag + reporting evidence \\
NIST AI RMF & MANAGE 1 (Monitoring) & Ongoing risk monitoring & Monitoring evidence in reports \\
NIST AI RMF & MANAGE 2 (Response) & Risk response actions taken & Remediation / correction keywords \\
NIST AI RMF & MANAGE 3 (Communication) & Transparency and disclosure & Disclosure / transparency keywords \\
NIST AI RMF & MANAGE 4 (Decommission) & System-withdrawal actions & Recall / suspension / shutdown keywords \\
NIST AI RMF & GOVERN 4 (Accountability) & Accountability structures & Responsibility / investigation keywords \\
GDPR & Art. 22 (Automated Decisions) & Human-oversight safeguards & HITL / explainability keywords \\
GDPR & Arts. 33--34 (Breach Notification) & Authority / subject notification & Notification / reporting keywords \\
\bottomrule
\end{tabular}
\end{table}

\subsection*{3.3 Analytical strategy}
The empirical analysis proceeds in three steps mapped to the research questions. To answer RQ1 and RQ2 we compute framework-level and provision-level compliance rates for the full sample and disaggregate by risk category, sector, and country. We interpret ``insufficient evidence'' separately from non-compliance because missing public documentation can reflect either an actual governance failure or a reporting limitation; analytically, however, both conditions weaken public accountability. To answer RQ3 we compare compliance rates conditional on incident detection source, report effect sizes, and test the EU AI Act comparison using Fisher's exact test. To answer RQ4 we translate the observed failure patterns into the PAGCF, treating the framework as a design implication derived from the empirical gaps rather than as an independently validated intervention.

\subsection*{3.4 Ethics and reproducibility}
The AIID is publicly released under a CC BY 4.0 licence and contains no personally identifiable information. The coding scripts, keyword dictionaries, and reproducibility notes will be released with the camera-ready version of the paper.

PART I: CROSS-REGULATORY GOVERNANCE GAP ANALYSIS

\section*{4. Results: Cross-Regulatory Governance Gaps}
\subsection*{4.1 Framework-level compliance}
Figure 1 summarizes compliance patterns across the three frameworks. The aggregate picture is not one of uniform failure; rather, each framework leaves a different empirical trace. The EU AI Act exposes a monitoring deficit, NIST captures a pattern of post-hoc remediation, and the GDPR reveals an evidentiary deficit in which the public record often does not disclose whether data-protection safeguards were undertaken.

\begin{figure}[t]
\centering
\includegraphics[width=0.9\linewidth]{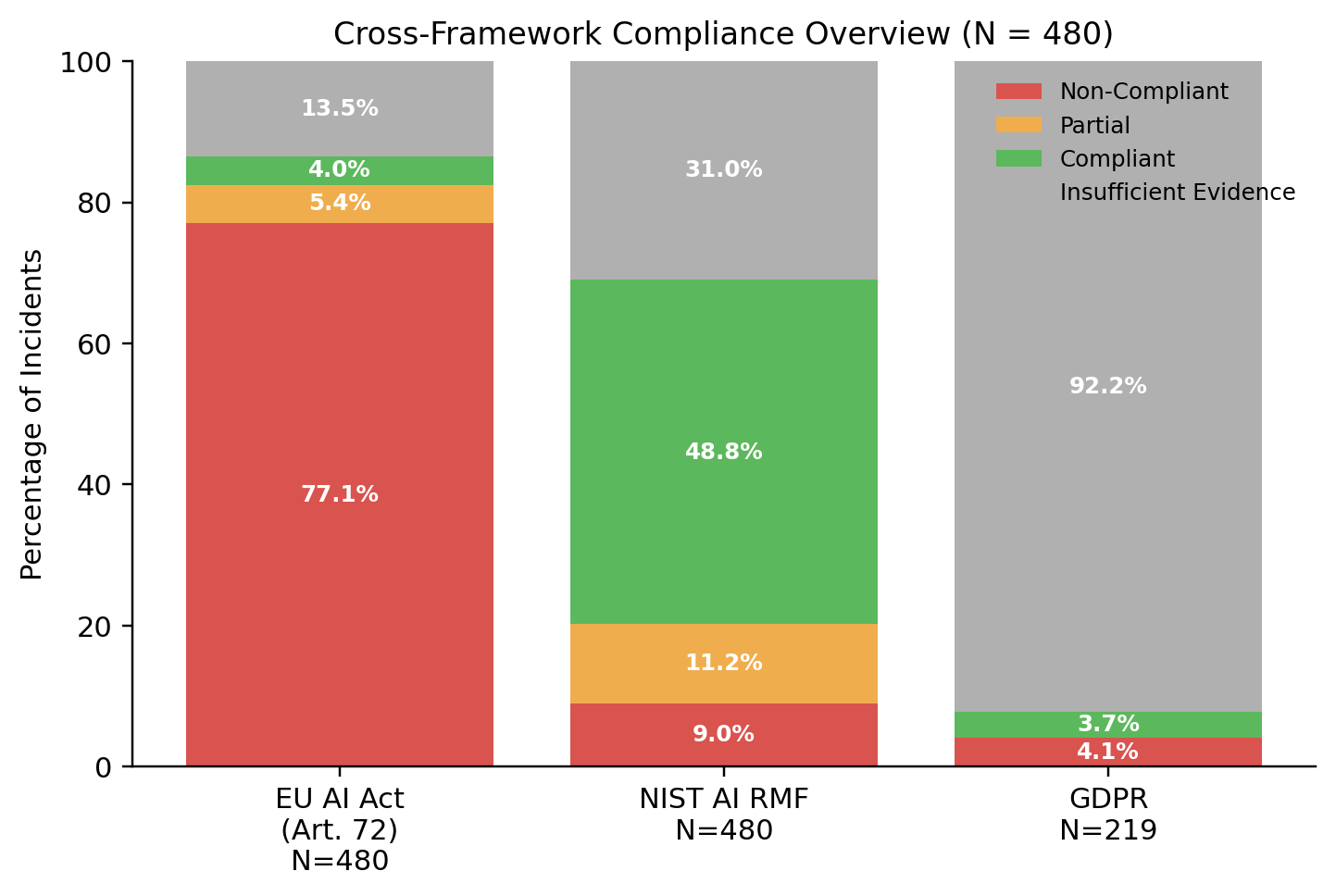}
\caption{Cross-framework compliance overview across 480 AI incidents. The GDPR bar reflects the n = 219 subset of incidents to which the regulation applies.}
\end{figure}

The EU AI Act exhibits the highest rate of non-compliance: 77.1\% of incidents show no evidence of the post-market monitoring required under Article 72. This result is especially important because Article 72 is not a peripheral documentation requirement; it is the mechanism through which high-risk systems are expected to remain governable after deployment. The NIST AI RMF shows a more mixed pattern; 48.8\% of incidents show high alignment, driven largely by MANAGE 4 (decommissioning). In many of these cases systems were recalled or suspended, but typically only after harm had occurred and in response to external pressure. NIST alignment therefore often reflects remedial action rather than continuous risk management. Among the 219 incidents to which the GDPR applies, only 7.8\% are conclusively classified (4.1\% non-compliant, 3.7\% compliant); 92.2\% lack sufficient documented evidence to make a definitive assessment. This large ``insufficient evidence'' band is itself a governance signal, because public accountability depends on precisely the kind of documentation that DPIA and Art. 22 provisions are intended to produce.

\subsection*{4.2 Provision-level compliance}
Figure 2 disaggregates compliance by provision. Post-market monitoring under EU AI Act Art. 72 shows the highest non-compliance rate at 77.1\%, followed by serious-incident reporting under Art. 73 (82.9\% of applicable incidents show no reporting evidence). In contrast, NIST MANAGE 4 (decommissioning) shows the strongest positive signal: 57.5\% of incidents involved some form of recall, suspension, or system removal. However, these actions were largely reactive.

\begin{figure}[t]
\centering
\includegraphics[width=0.9\linewidth]{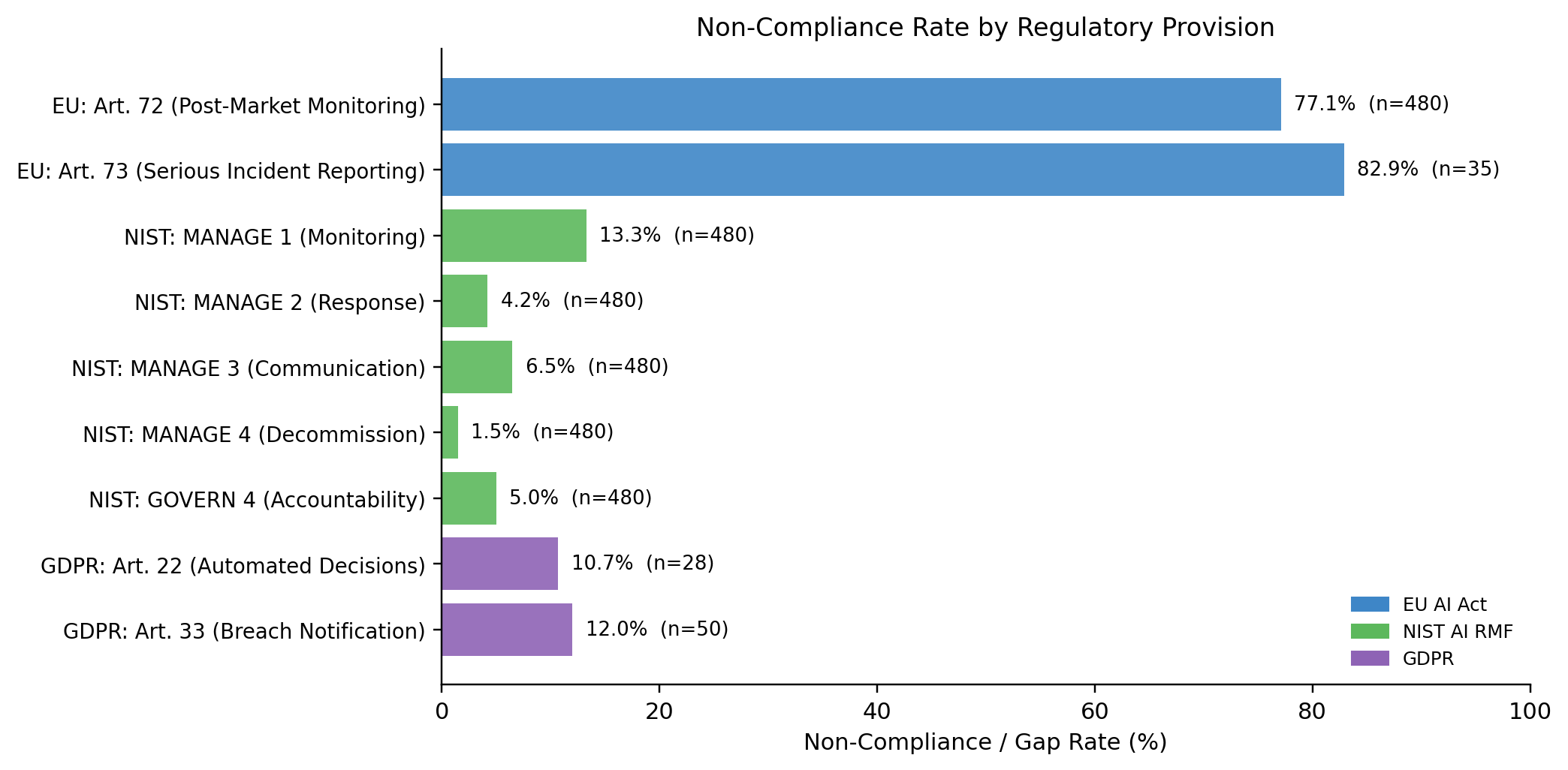}
\caption{Non-compliance rate by regulatory provision. Bar colour denotes framework; denominators vary by provision applicability.}
\end{figure}

Proactive mechanisms fare worst. Continuous monitoring under NIST MANAGE 1 is present in only 9.4\% of incidents, and evidence of a Data-Protection Impact Assessment under GDPR Art. 35 appears in only 0.4\%. The provision-level results therefore distinguish two forms of governance failure. Some mechanisms are absent even after harm becomes public; others may exist internally but leave no observable accountability trail. Both conditions are problematic for post-deployment governance, whose purpose is to make deployed systems continuously inspectable rather than merely remediable after public controversy.

\subsection*{4.3 Are governance failures framework-specific or systemic?}
A central question for regulatory design is whether governance gaps are specific to individual frameworks or reflect broader systemic weaknesses. Of the 480 incidents analysed, 47 (9.8\%) were flagged as non-compliant under two or more frameworks simultaneously. The pairwise Jaccard overlap between frameworks is consistently low but non-zero (2.2--11.3\%). This pattern has two interpretations. The low overlap indicates that each framework captures a partly distinct accountability dimension; the non-zero overlap indicates that some incidents traverse multiple accountability boundaries without triggering the expected controls. Regulatory plurality therefore does not function as a safety net by default. Instead, the results are consistent with the accountability supply chain argument advanced by Cobbe, Veale, and Singh [3]: failures arise from coordination gaps between actors, documentation systems, and oversight institutions, not simply from a lack of formal frameworks.

\section*{5. The Internal-Monitoring Effect}
The most consequential finding of the incident analysis is the large compliance gap between incidents detected through internal monitoring and those identified externally. Detection source matters because it indicates where accountability first becomes operational: inside the deploying organisation, or only after external actors have surfaced the harm. Of the 480 incidents analysed, only 24 (5.0\%) were detected through internal monitoring mechanisms; the remaining 456 (95.0\%) were identified by journalists, users, researchers, regulators, or through legal proceedings. Compliance outcomes differ sharply between the two groups (Figure 3). Under EU AI Act Art. 72, 87.5\% of internally detected incidents were classified as compliant or partially compliant, versus 5.3\% of externally detected incidents a 16.5$\times$ (rounded to 17$\times$) difference. A similar pattern appears under the NIST AI RMF: 95.8\% of internally detected incidents show high or moderate alignment versus 58.1\% of externally detected incidents (1.65$\times$). A two-sided Fisher's exact test on the EU AI Act comparison yields p $<$ 0.001, indicating that the difference is extremely unlikely under the null of independence between detection source and compliance outcome.

\begin{figure}[t]
\centering
\includegraphics[width=0.9\linewidth]{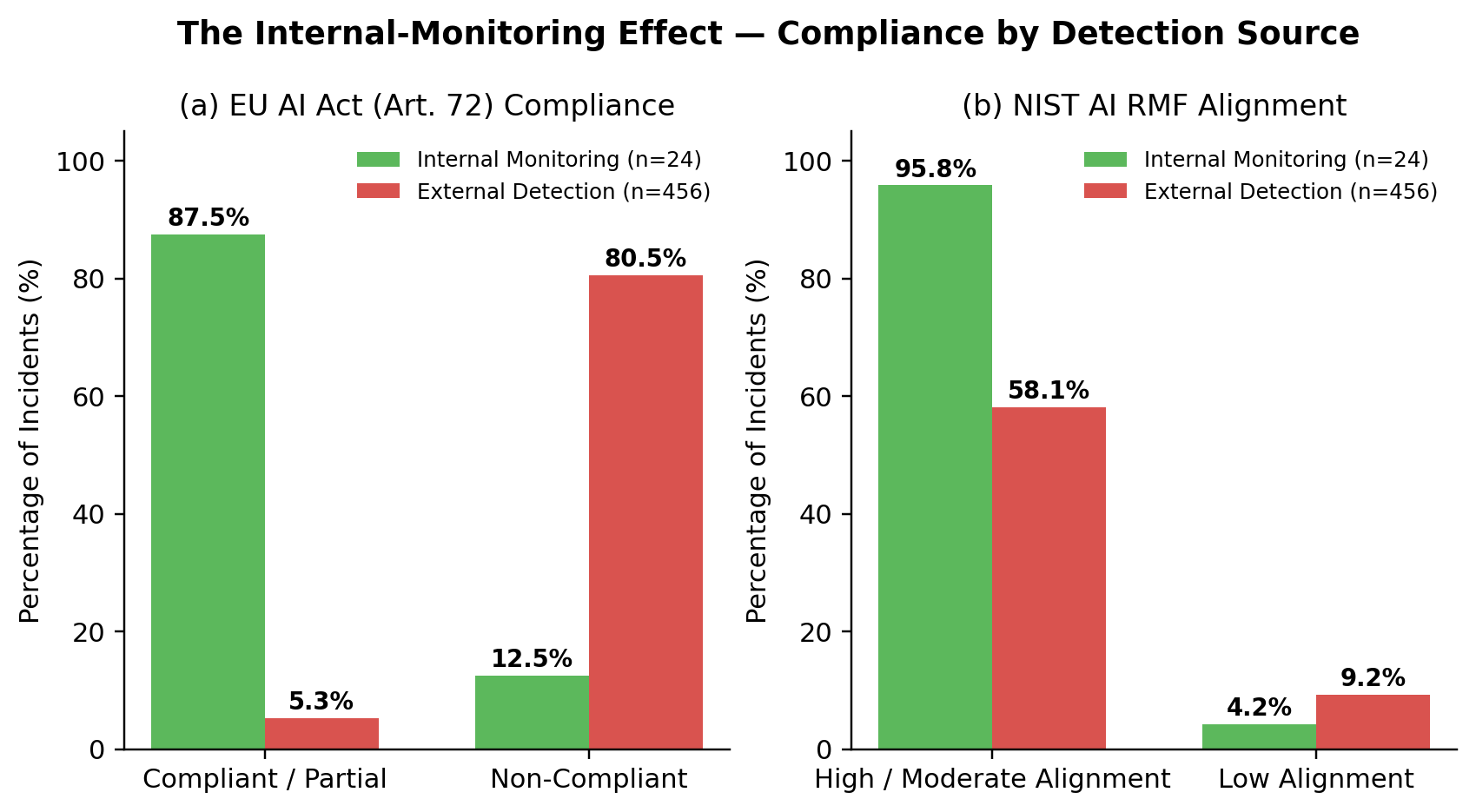}
\caption{Compliance rates for internally detected incidents (n = 24) versus externally detected incidents (n = 456). Internal monitoring is associated with dramatically higher compliance under both frameworks.}
\end{figure}

We stress two caveats. First, the internally detected sample is small (n = 24) and organisations that implement internal monitoring may already be more mature in their governance practices, so the observed compliance advantage may reflect broader organisational capacity rather than monitoring alone a classic selection-on-observables concern [21]. Second, the AIID probably underrepresents incidents that are detected and resolved internally before becoming public, biasing the internally detected sample toward comparatively well-governed cases. Even so, the magnitude of the difference and its consistency across two independent frameworks suggests that internal monitoring is, at a minimum, a strong indicator of better compliance and plausibly a direct contributor to it. Substantively, the result shifts the governance question from whether organisations possess formal policies to whether they possess sensing mechanisms capable of discovering failures before outsiders do.

\section*{6. The Proactive AI Governance Compliance Framework (PAGCF)}
Every governance element that the frameworks assume to be in place before an incident is absent in more than 90\% of cases (Table 2). The central design implication is therefore not simply that organisations should respond faster after incidents occur. Rather, they require governance infrastructure that creates evidence, escalation pathways, and accountability assignments before harm becomes externally visible.

\begin{table}[t]
\centering
\footnotesize
\caption{Pre-incident governance gaps and their regulatory basis.}
\begin{tabular}{@{}p{0.42\textwidth} p{0.20\textwidth} p{0.28\textwidth}@{}}
\toprule
Pre-Incident Governance Element & Gap Rate & Source Provision \\
\midrule
Pre-Deployment Impact Assessment (DPIA) & 99.6\% absent & GDPR Art. 35 \\
Internal Detection / Monitoring & 95.0\% externally detected & NIST MANAGE 1 \\
Continuous Monitoring Systems & 90.6\% absent & EU AI Act Art. 72; NIST MANAGE 1 \\
Accountability Structures & 91.5\% absent & NIST GOVERN 4 \\
Incident Response Plans & 92.7\% absent & NIST MANAGE 2 \\
Communication Protocols & 90.6\% absent & NIST MANAGE 3 \\
\bottomrule
\end{tabular}
\end{table}

\subsection*{6.1 Architecture: four sequential phases}
The PAGCF (Figure 4) consists of four sequential phases covering the full AI system lifecycle. Each phase is mapped to specific provisions of the EU AI Act, NIST AI RMF, and GDPR and is grounded in the governance gaps identified above. The framework is deliberately provision-oriented: its purpose is to translate regulatory requirements into auditable organisational routines rather than to introduce another high-level ethics checklist.

\begin{figure}[t]
\centering
\includegraphics[width=0.9\linewidth]{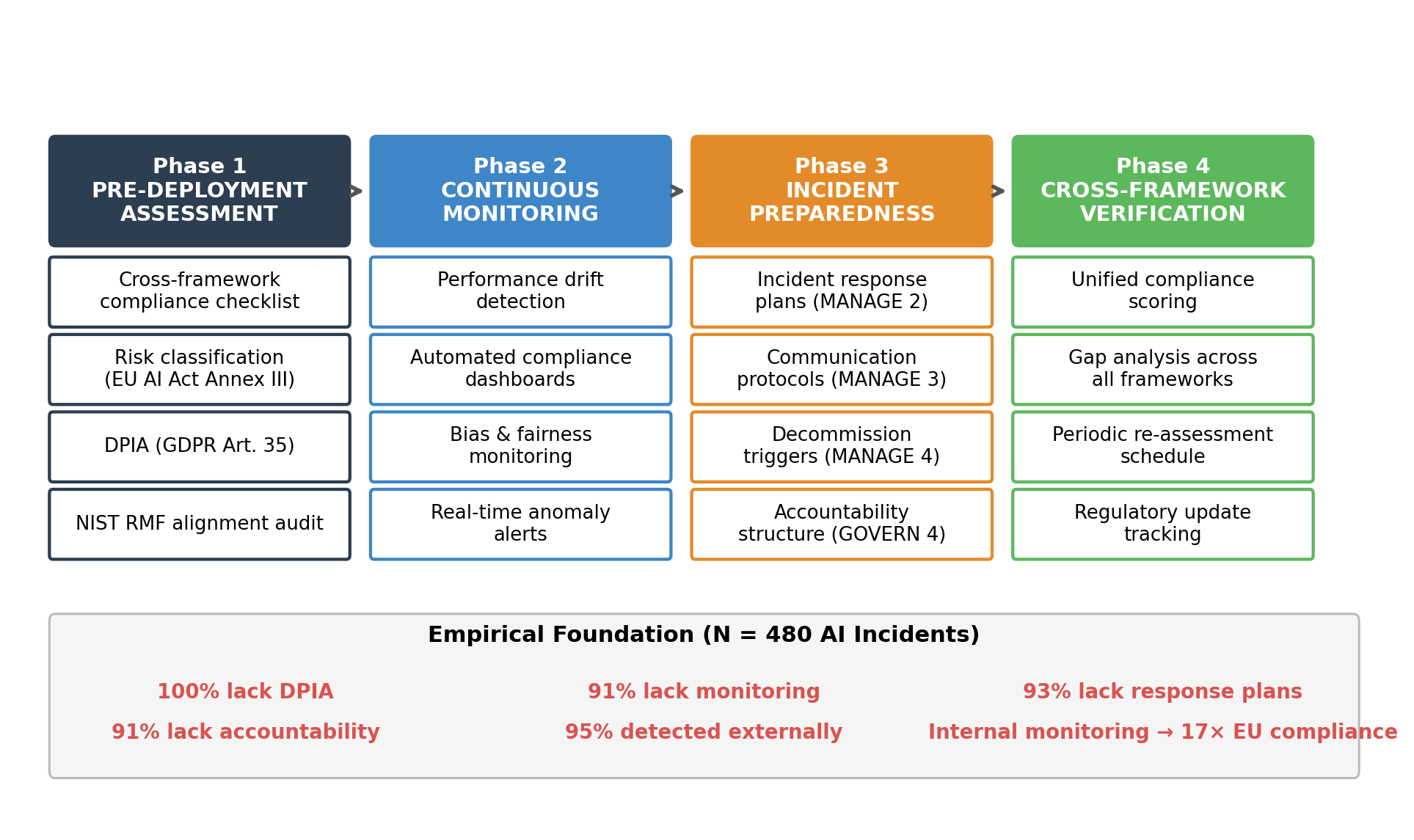}
\caption{The PAGCF four-phase architecture. Each phase is mapped to specific regulatory provisions and supported by empirical evidence from the incident sample.}
\end{figure}

Phase 1: Pre-Deployment Assessment. Phase 1 represents the most critical point of intervention. Because 99.6\% of incidents show no evidence of a DPIA and 90.6\% lack monitoring systems at the time of the incident, this phase requires a cross-framework compliance checklist that unifies key requirements from EU AI Act Art. 9 (risk management), the NIST GOVERN functions, and GDPR Art. 35. It also requires risk classification using the EU AI Act Annex III categories, motivated by the 60--87\% variation in non-compliance rates across risk categories (Figure 5); a Data-Protection Impact Assessment for any AI system processing personal data; and a pre-deployment alignment audit against NIST AI RMF MANAGE 1--4 and GOVERN 4 to establish a baseline for later monitoring.

Phase 2: Continuous Monitoring. Phase 2 is the highest-impact intervention and is directly supported by the central finding in Section 5. It comprises four components: automated performance monitoring (fairness metrics, output quality, addressing Art. 72); drift-detection systems (data drift, concept drift, performance degradation); real-time compliance dashboards mapping system status to specific regulatory provisions; and anomaly alerting tied to pre-deployment baselines that escalates into the Phase 3 preparedness protocols. The purpose of these components is not only technical reliability, but institutional observability: they make it possible to know when governance obligations have been triggered. This mirrors the internal-audit architecture proposed by Raji et al. [20] and the layered auditing model in Mökander and Schuett [16].

Phase 3: Incident Preparedness. Because 92.7\% of incidents lack response plans, 90.6\% lack communication protocols, and 91.5\% lack accountability structures, Phase 3 requires four governance mechanisms to be established before incidents occur: predefined incident response plans (aligned with MANAGE 2); communication protocols (aligned with MANAGE 3 and GDPR Art. 33's 72-hour clock); decommission triggers specified in advance (MANAGE 4); and clear accountability assignments (GOVERN 4).

Phase 4: Cross-Framework Verification. Phase 4 responds directly to the finding that 9.8\% of incidents fail under two or more frameworks simultaneously. It requires unified compliance scoring across all applicable frameworks, periodic gap analysis, systematic tracking of regulatory updates, and integration of lessons learned back into the Phase 1 assessment criteria creating the feedback loop that current governance practice conspicuously lacks.

\subsection*{6.2 Risk-stratified governance tiers}
AI systems do not require uniform governance intensity. The observed variation in non-compliance across risk categories (Figure 5) motivates a tiered approach summarized in Table 3.

\begin{figure}[t]
\centering
\includegraphics[width=0.9\linewidth]{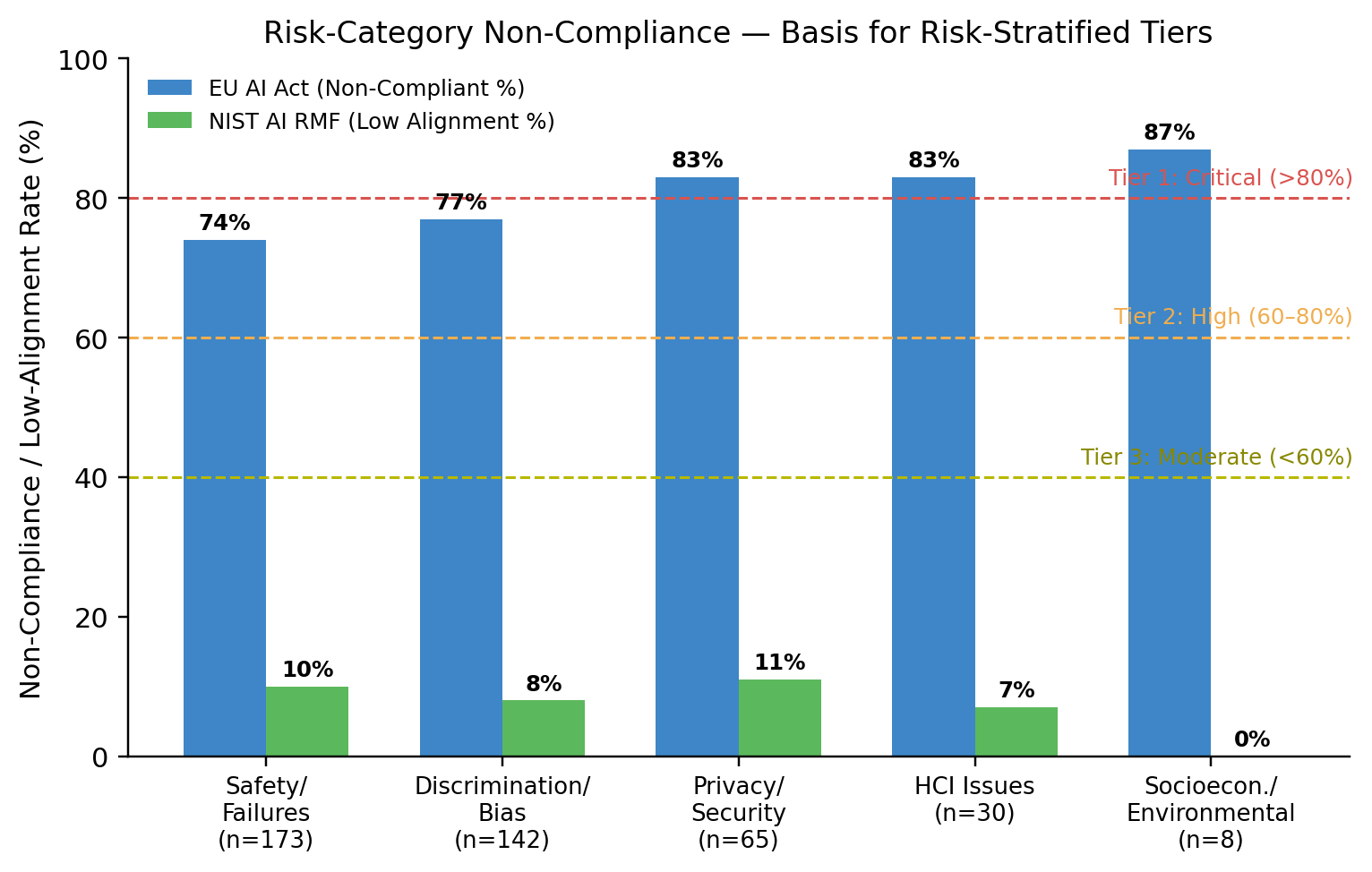}
\caption{Non-compliance rates by risk category with proposed governance-tier thresholds.}
\end{figure}

\begin{table}[t]
\centering
\footnotesize
\caption{PAGCF risk-stratified governance tiers.}
\begin{tabular}{@{}p{0.18\textwidth} p{0.24\textwidth} p{0.27\textwidth} p{0.23\textwidth}@{}}
\toprule
Governance Tier & Risk Profile & PAGCF Requirement & Assessment Frequency \\
\midrule
Tier 1: Critical & Safety/Failures, Discrimination/Bias & All four phases mandatory; continuous monitoring; quarterly external audits & Continuous + Quarterly \\
Tier 2: High & Privacy/Security, Socioeconomic/Environmental & Phases 1--3 mandatory; monitoring recommended; semi-annual audits & Monthly + Semi-Annual \\
Tier 3: Moderate & HCI Issues, low-risk applications & Phase 1 mandatory; monitoring recommended; annual audits & Quarterly + Annual \\
\bottomrule
\end{tabular}
\end{table}

\subsection*{6.3 Projected impact}
We cannot run a controlled experiment. However, the 24 internally monitored incidents serve as a natural proxy for what proactive governance looks like in the wild. Figure 6 compares current governance state (all 480 incidents) with this proxy subset. The point estimates suggest that EU AI Act compliance could rise from below 10\% to above 85\%, and NIST AI RMF alignment from 60\% to over 95\%. Selection bias cannot be ruled out, and the proxy should not be interpreted as a causal effect estimate. The projection is best read as a prioritisation argument: if organisations can implement only one governance intervention first, the evidence points to internal monitoring as the intervention most closely associated with accountable post-deployment behaviour.

\begin{figure}[t]
\centering
\includegraphics[width=0.9\linewidth]{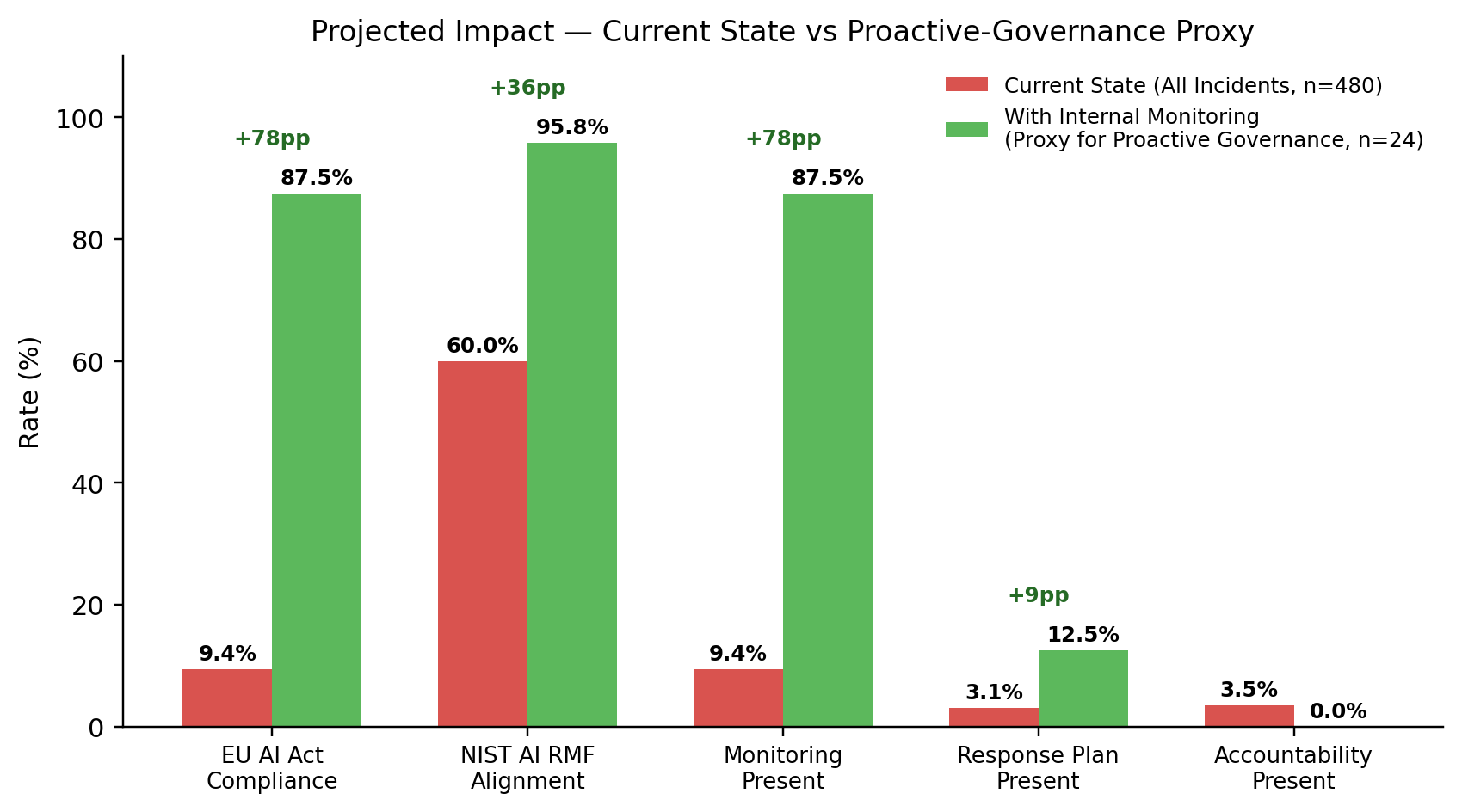}
\caption{Current governance state (all incidents) versus internally monitored incidents (proxy for proactive governance). Improvements are substantial across all dimensions but should be interpreted as associational rather than causal.}
\end{figure}

\subsection*{6.4 The governance-lifecycle shift}
The PAGCF fundamentally shifts when governance effort is applied (Figure 7). Current practice is largely reactive: 95\% of incidents are detected externally, so oversight functions as a post-hoc response rather than a preventive mechanism. The PAGCF reorients governance toward pre-deployment assessment and continuous monitoring, where the evidence indicates the greatest leverage. In conceptual terms, the shift is from incident response as reputational damage control to governance as a continuous evidentiary system: one that records baselines, detects deviations, assigns responsibility, and creates the documentation needed for regulatory review.

\begin{figure}[t]
\centering
\includegraphics[width=0.9\linewidth]{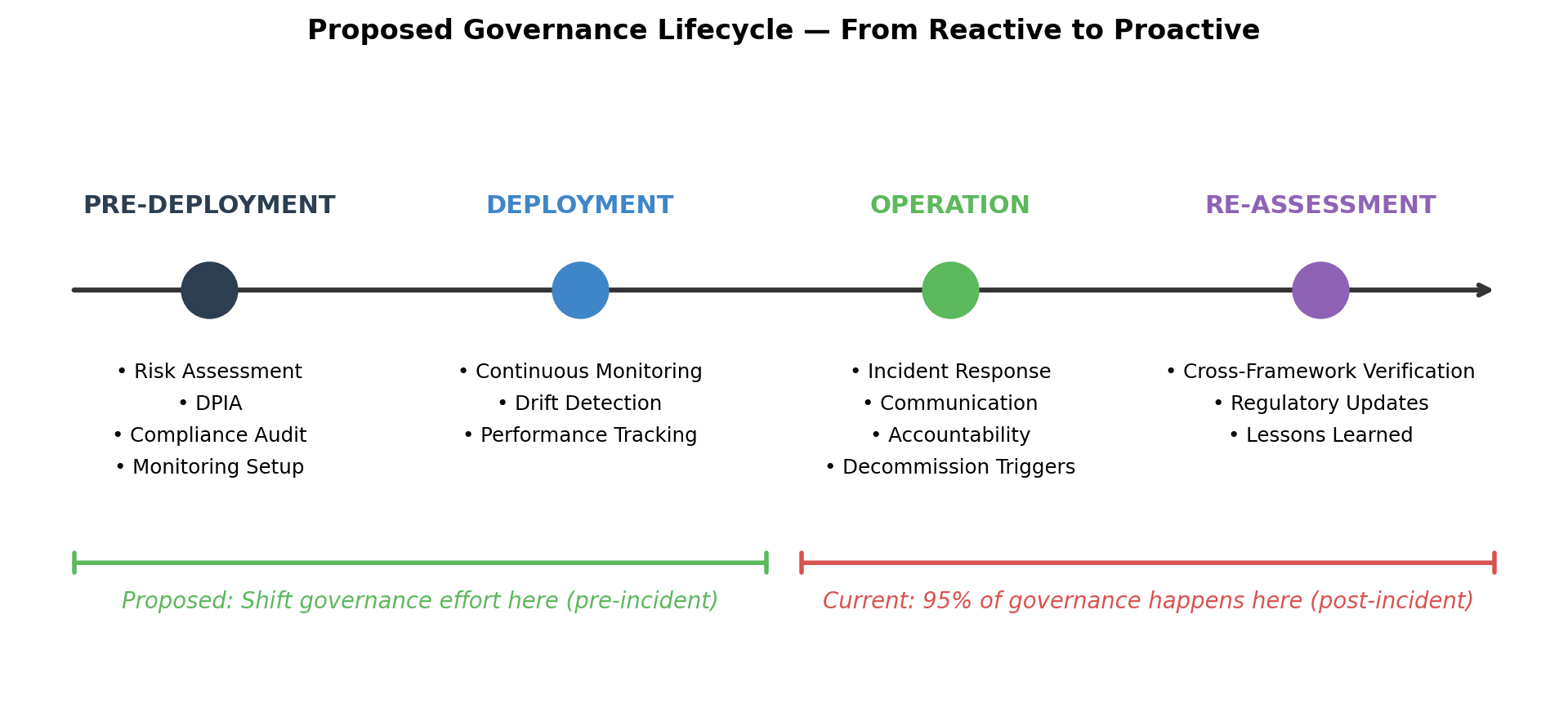}
\caption{Proposed governance-lifecycle shift. Green bracket: proposed allocation of governance effort (pre-incident). Red bracket: current allocation, where 95\% of governance activity occurs post-incident.}
\end{figure}

\section*{7. Discussion}
Three overarching findings emerge from the empirical analysis. First, governance failures are pervasive across all frameworks: no regime shows consistently effective post-deployment performance. The EU AI Act records 77.1\% non-compliance on post-market monitoring, and even the GDPR the regime with the most mature enforcement infrastructure leaves 92.2\% of applicable incidents evidentially indeterminate. Second, governance gaps are systemic rather than framework-specific: 9.8\% of incidents are non-compliant under two or more frameworks simultaneously, so the core problem is less about choosing the right framework than about ensuring that governance mechanisms function in operational settings. Third, internal monitoring appears to be a major compliance catalyst: internally detected incidents show 87.5\% EU compliance versus 5.3\% for externally detected incidents, suggesting that proactive monitoring may be the highest-leverage available intervention. Taken together, these findings imply that post-deployment accountability is primarily an implementation problem: the formal architecture exists, but the organisational routines needed to make it visible and enforceable are often missing.

These findings speak to three broader debates. First, they call into question the assumption that a proliferation of regulatory instruments necessarily leads to better governance outcomes an empirical version of the governance-theatre critique advanced by Hagendorff [6] and Ali and Christin [1], in which regulatory artefacts multiply without corresponding implementation. Second, the comparatively stronger GDPR outcomes on privacy-related incidents are consistent with the argument that enforcement capacity financial penalties, established supervisory authorities materially matters [25, 24]. Yet the GDPR still shows important gaps in AI-specific contexts, indicating that enforcement alone is insufficient without technical standards tailored to AI systems. Third, the predominance of external detection (95\% of cases) challenges a core design assumption underlying all three frameworks: that deployers will reliably monitor and report their own harms. In practice, journalists, civil-society actors, researchers, and affected individuals function as the primary mechanisms of accountability a pattern echoing the de facto distributed audit ecology described by Raji et al. [20] and Metcalf et al. [12].

Policy implications. Four implications follow. First, policymakers should prioritise enforcement capacity over the creation of additional frameworks: binding obligations backed by penalties appear to outperform voluntary approaches, but only where obligations are specific enough to generate auditable records. Second, monitoring requirements should be supported by technical standards, certification mechanisms, and evidence-preservation duties the absence of documentation is itself the largest observed governance gap. Third, governance frameworks should assign a more explicit role to independent auditors, civil-society monitors, and affected communities rather than relying primarily on deployer self-governance. The empirical dominance of external detection shows that these actors are already performing core accountability work, even when regulatory frameworks do not fully recognise that role. Fourth, the overlap in non-compliance across frameworks points to the need for cross-framework harmonisation; unified incident reporting across jurisdictions could reduce compliance burdens while strengthening overall coverage. The PAGCF offers a practical starting point: organisations should begin with Phase 2 (continuous monitoring) as the highest-impact intervention and then extend implementation across the remaining phases.

\section*{8. Limitations}
Data source. The AIID captures publicly documented AI failures and therefore probably underrepresents incidents that are detected and resolved internally before becoming public. This limitation biases our sample toward externally detected cases and may inflate the observed external-detection majority; it also biases the internally detected subset toward comparatively well-governed organisations.

Proxy for proactive governance. The proxy for proactive governance is based on only n = 24 internally detected incidents. Although the effect size is substantial and highly statistically significant, a larger sample and, ideally, a matched design would strengthen the causal claim.

\section*{9. Conclusion}
This paper evaluated 480 real-world AI incidents against three major governance frameworks the EU AI Act, the NIST AI Risk Management Framework, and the GDPR to provide a provision-level empirical assessment of AI-governance effectiveness after deployment. The findings show that governance failures are pervasive, systemic, and only weakly differentiated by framework. More than 93\% of incidents lack basic pre-incident governance elements: monitoring systems, response plans, accountability structures, and communication protocols. This analysis provides the empirical basis for the Proactive AI Governance Compliance Framework (PAGCF), a four-phase lifecycle-deployment assessment, continuous monitoring, incident preparedness, and cross-framework verification with risk-stratified governance tiers calibrated to the observed variation in non-compliance across risk categories.

The results suggest that the shift from reactive to proactive AI governance is not merely desirable but potentially the highest-leverage intervention available to regulators, deployers, and policymakers seeking to reduce AI harms. For post-deployment accountability to become more than a formal legal promise, organisations must be able to produce timely evidence of monitoring, escalation, remediation, and oversight. Future research should (i) extend the analysis to additional regulatory regimes and to sector-specific supervisors, (ii) benchmark the automated coding scheme against expert-adjudicated ground truth, (iii) track outcomes longitudinally as the EU AI Act enters full force, and (iv) pilot the PAGCF in organisational settings to evaluate whether the projected compliance improvements can be realised in practice.

\section*{Data \& Code Availability}
The AI Incident Database is publicly available at \url{https://incidentdatabase.ai}. Coding scripts, keyword dictionaries, and figure-regeneration code will be released with the camera-ready version of the paper.

\section*{References}
\hangindent=1.5em\hangafter=1 [1] Ali, S. J., \& Christin, A. (2023). Walking the walk of AI ethics: Organizational challenges and the individualization of risk among ethics entrepreneurs. In Proceedings of the 2023 ACM Conference on Fairness, Accountability, and Transparency (FAccT '23) (pp. 217--226). ACM. \url{https://doi.org/10.1145/3593013.3593990}\par
\hangindent=1.5em\hangafter=1 [2] Berente, N., Gu, B., Recker, J., \& Santhanam, R. (2021). Managing artificial intelligence. MIS Quarterly, 45(3), 1433--1450. \url{https://doi.org/10.25300/MISQ/2021/16274}\par
\hangindent=1.5em\hangafter=1 [3] Cobbe, J., Veale, M., \& Singh, J. (2023). Understanding accountability in algorithmic supply chains. In Proceedings of the 2023 ACM Conference on Fairness, Accountability, and Transparency (FAccT '23) (pp. 1186--1197). ACM. \url{https://doi.org/10.1145/3593013.3594073}\par
\hangindent=1.5em\hangafter=1 [4] European Parliament \& Council of the European Union. (2016). Regulation (EU) 2016/679 on the protection of natural persons with regard to the processing of personal data and on the free movement of such data (General Data Protection Regulation). Official Journal of the European Union, L 119, 1--88.\par
\hangindent=1.5em\hangafter=1 [5] European Parliament \& Council of the European Union. (2024). Regulation (EU) 2024/1689 laying down harmonised rules on artificial intelligence (Artificial Intelligence Act). Official Journal of the European Union, L, 2024/1689.\par
\hangindent=1.5em\hangafter=1 [6] Hagendorff, T. (2020). The ethics of AI ethics: An evaluation of guidelines. Minds and Machines, 30(1), 99--120. \url{https://doi.org/10.1007/s11023-020-09517-8}\par
\hangindent=1.5em\hangafter=1 [7] Jobin, A., Ienca, M., \& Vayena, E. (2019). The global landscape of AI ethics guidelines. Nature Machine Intelligence, 1(9), 389--399. \url{https://doi.org/10.1038/s42256-019-0088-2}\par
\hangindent=1.5em\hangafter=1 [8] Kroll, J. A., Huey, J., Barocas, S., Felten, E. W., Reidenberg, J. R., Robinson, D. G., \& Yu, H. (2017). Accountable algorithms. University of Pennsylvania Law Review, 165(3), 633--705.\par
\hangindent=1.5em\hangafter=1 [9] Laux, J., Wachter, S., \& Mittelstadt, B. (2024). Trustworthy artificial intelligence and the European Union AI Act: On the conflation of trust and acceptability. Regulation \& Governance, 18(1), 3--32. \url{https://doi.org/10.1111/rego.12512}\par
\hangindent=1.5em\hangafter=1 [10] Mahler, T. (2022). Between risk management and proportionality: The risk-based approach in the EU's Artificial Intelligence Act proposal. In Nordic Yearbook of Law and Informatics 2020--2021 (pp. 247--270). \url{https://doi.org/10.53292/208f5901.38a67238}\par
\hangindent=1.5em\hangafter=1 [11] McGregor, S. (2021). Preventing repeated real-world AI failures by cataloging incidents: The AI Incident Database. Proceedings of the AAAI Conference on Artificial Intelligence, 35(17), 15458--15463. \url{https://doi.org/10.1609/aaai.v35i17.17817}\par
\hangindent=1.5em\hangafter=1 [12] Metcalf, J., Moss, E., Watkins, E. A., Singh, R., \& Elish, M. C. (2021). Algorithmic impact assessments and accountability: The co-construction of impacts. In Proceedings of the 2021 ACM Conference on Fairness, Accountability, and Transparency (FAccT '21) (pp. 735--746). ACM. \url{https://doi.org/10.1145/3442188.3445935}\par
\hangindent=1.5em\hangafter=1 [13] Mitchell, M., Wu, S., Zaldivar, A., Barnes, P., Vasserman, L., Hutchinson, B., Spitzer, E., Raji, I. D., \& Gebru, T. (2019). Model cards for model reporting. In Proceedings of the Conference on Fairness, Accountability, and Transparency (FAT* '19) (pp. 220--229). ACM. \url{https://doi.org/10.1145/3287560.3287596}\par
\hangindent=1.5em\hangafter=1 [14] Mittelstadt, B. D., Allo, P., Taddeo, M., Wachter, S., \& Floridi, L. (2016). The ethics of algorithms: Mapping the debate. Big Data \& Society, 3(2), 1--21. \url{https://doi.org/10.1177/2053951716679679}\par
\hangindent=1.5em\hangafter=1 [15] Mökander, J., Morley, J., Taddeo, M., \& Floridi, L. (2021). Ethics-based auditing of automated decision-making systems: Nature, scope, and limitations. Science and Engineering Ethics, 27(4), Article 44. \url{https://doi.org/10.1007/s11948-021-00319-4}\par
\hangindent=1.5em\hangafter=1 [16] Mökander, J., Schuett, J., Kirk, H. R., \& Floridi, L. (2023). Auditing large language models: A three-layered approach. AI and Ethics. \url{https://doi.org/10.1007/s43681-023-00289-2}\par
\hangindent=1.5em\hangafter=1 [17] Nach, H. (2025). Governance of artificial intelligence: An analysis of incidents from the AI Incident Database. In Proceedings of the 2025 IEEE International Conference on Technology Management, Operations and Decisions (ICTMOD) (pp. 1--6). IEEE.\par
\hangindent=1.5em\hangafter=1 [18] National Institute of Standards and Technology. (2023). Artificial Intelligence Risk Management Framework (AI RMF 1.0) (NIST AI 100-1). U.S. Department of Commerce. \url{https://doi.org/10.6028/NIST.AI.100-1}\par
\hangindent=1.5em\hangafter=1 [19] Novelli, C., Taddeo, M., \& Floridi, L. (2023). Accountability in artificial intelligence: What it is and how it works. AI \& Society, 39, 1871--1882. \url{https://doi.org/10.1007/s00146-023-01635-y}\par
\hangindent=1.5em\hangafter=1 [20] Raji, I. D., Smart, A., White, R. N., Mitchell, M., Gebru, T., Hutchinson, B., Smith-Loud, J., Theron, D., \& Barnes, P. (2020). Closing the AI accountability gap: Defining an end-to-end framework for internal algorithmic auditing. In Proceedings of the 2020 Conference on Fairness, Accountability, and Transparency (FAT* '20) (pp. 33--44). ACM. \url{https://doi.org/10.1145/3351095.3372873}\par
\hangindent=1.5em\hangafter=1 [21] Rakova, B., Yang, J., Cramer, H., \& Chowdhury, R. (2021). Where responsible AI meets reality: Practitioner perspectives on enablers for shifting organizational practices. Proceedings of the ACM on Human-Computer Interaction, 5(CSCW1), Article 7. \url{https://doi.org/10.1145/3449081}\par
\hangindent=1.5em\hangafter=1 [22] Selbst, A. D., Boyd, D., Friedler, S. A., Venkatasubramanian, S., \& Vertesi, J. (2019). Fairness and abstraction in sociotechnical systems. In Proceedings of the Conference on Fairness, Accountability, and Transparency (FAT* '19) (pp. 59--68). ACM. \url{https://doi.org/10.1145/3287560.3287598}\par
\hangindent=1.5em\hangafter=1 [23] Smuha, N. A. (2021). From a `race to AI' to a `race to AI regulation': Regulatory competition for artificial intelligence. Law, Innovation and Technology, 13(1), 57--84. \url{https://doi.org/10.1080/17579961.2021.1898300}\par
\hangindent=1.5em\hangafter=1 [24] Veale, M., \& Zuiderveen Borgesius, F. (2021). Demystifying the Draft EU Artificial Intelligence Act. Computer Law Review International, 22(4), 97--112. \url{https://doi.org/10.9785/cri-2021-220402}\par
\hangindent=1.5em\hangafter=1 [25] Wachter, S., Mittelstadt, B., \& Floridi, L. (2017). Why a right to explanation of automated decision-making does not exist in the General Data Protection Regulation. International Data Privacy Law, 7(2), 76--99. \url{https://doi.org/10.1093/idpl/ipx005}\par
\hangindent=1.5em\hangafter=1 [26] Wei, M., \& Zhou, Z. (2023). AI ethics issues in practice: Evidence from AI Incident Database. In Proceedings of the 56th Hawaii International Conference on System Sciences (HICSS) (pp. 5765--5774). \url{https://doi.org/10.24251/HICSS.2023.700}\par
\hangindent=1.5em\hangafter=1 [27] Wieringa, M. (2020). What to account for when accounting for algorithms: A systematic literature review on algorithmic accountability. In Proceedings of the 2020 Conference on Fairness, Accountability, and Transparency (FAT* '20) (pp. 1--18). ACM. \url{https://doi.org/10.1145/3351095.3372833}\par
\section*{Appendix A: Additional Figures}
\begin{figure}[t]
\centering
\includegraphics[width=0.9\linewidth]{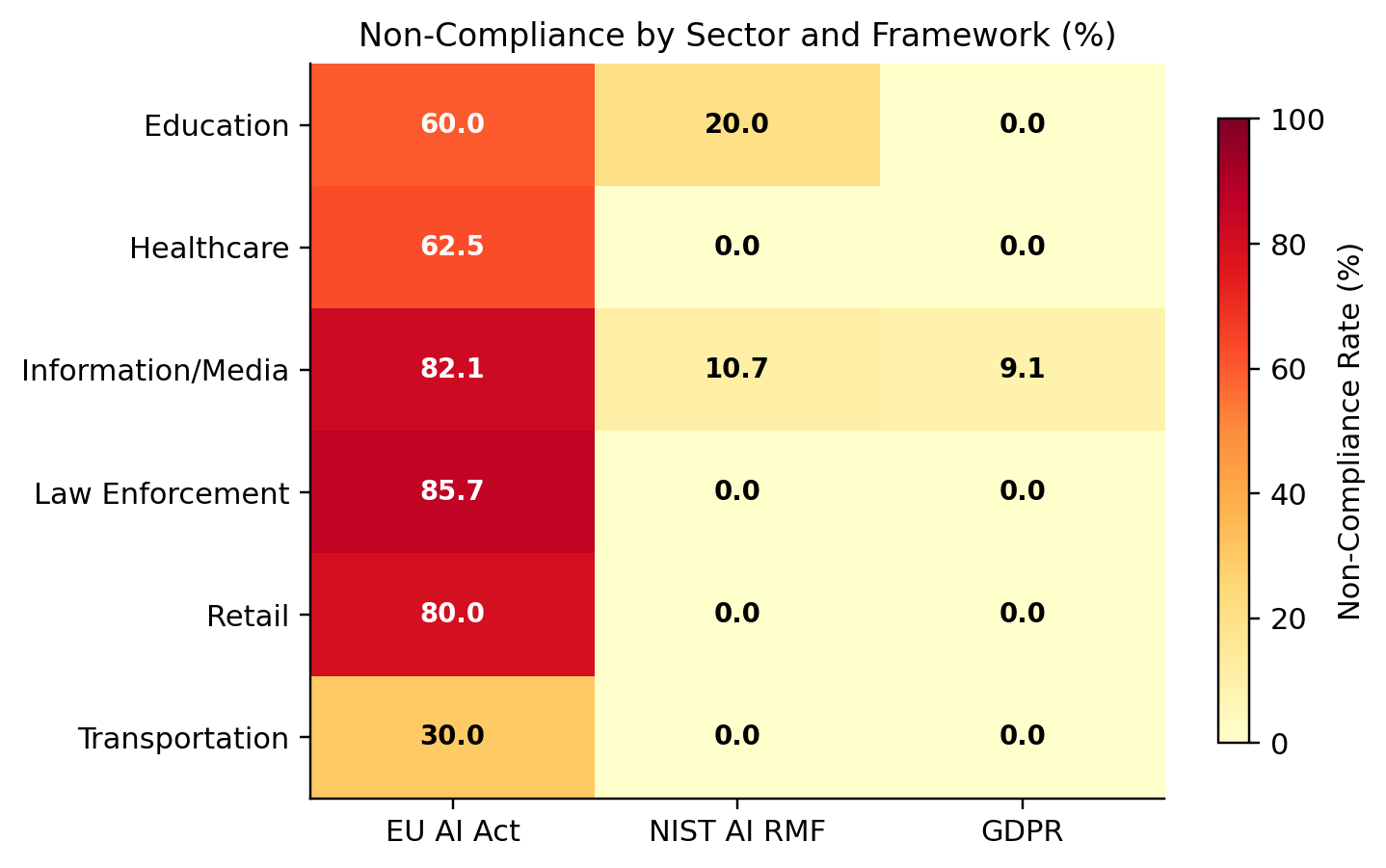}
\caption{Non-compliance by sector and framework.}
\end{figure}

\begin{figure}[t]
\centering
\includegraphics[width=0.9\linewidth]{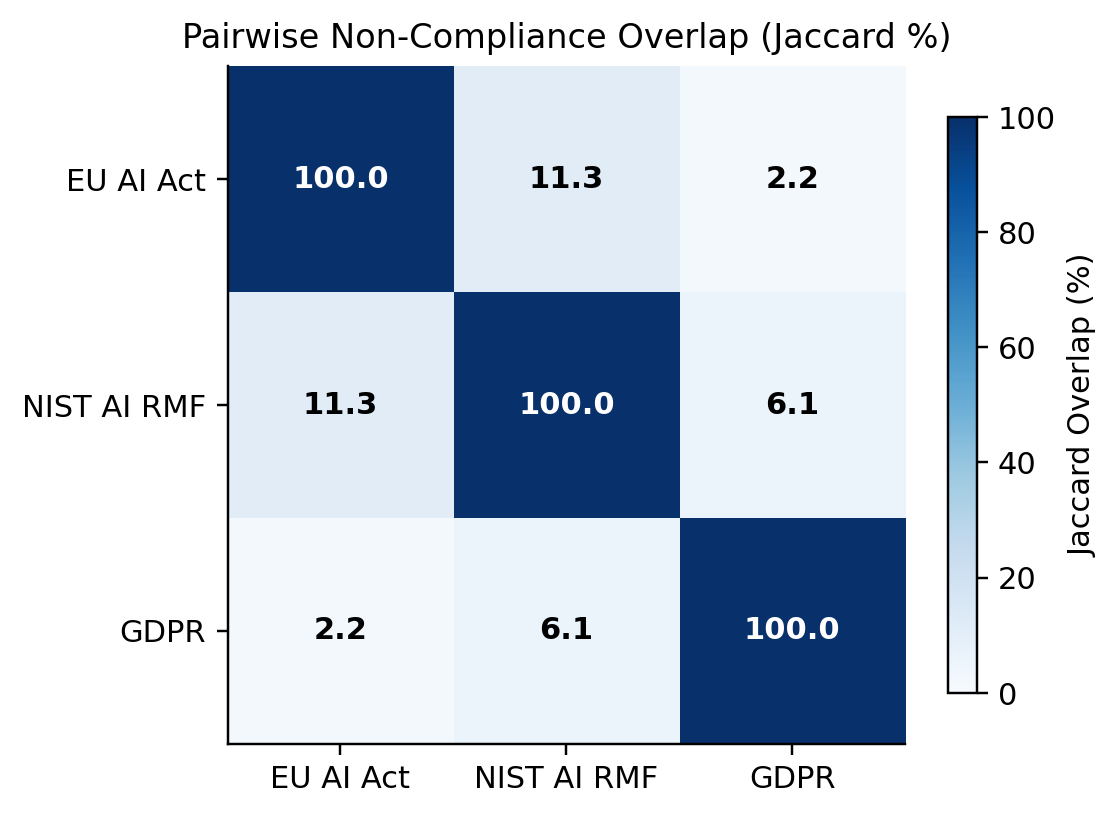}
\caption{Pairwise non-compliance overlap between frameworks (Jaccard \%).}
\end{figure}

\begin{figure}[t]
\centering
\includegraphics[width=0.9\linewidth]{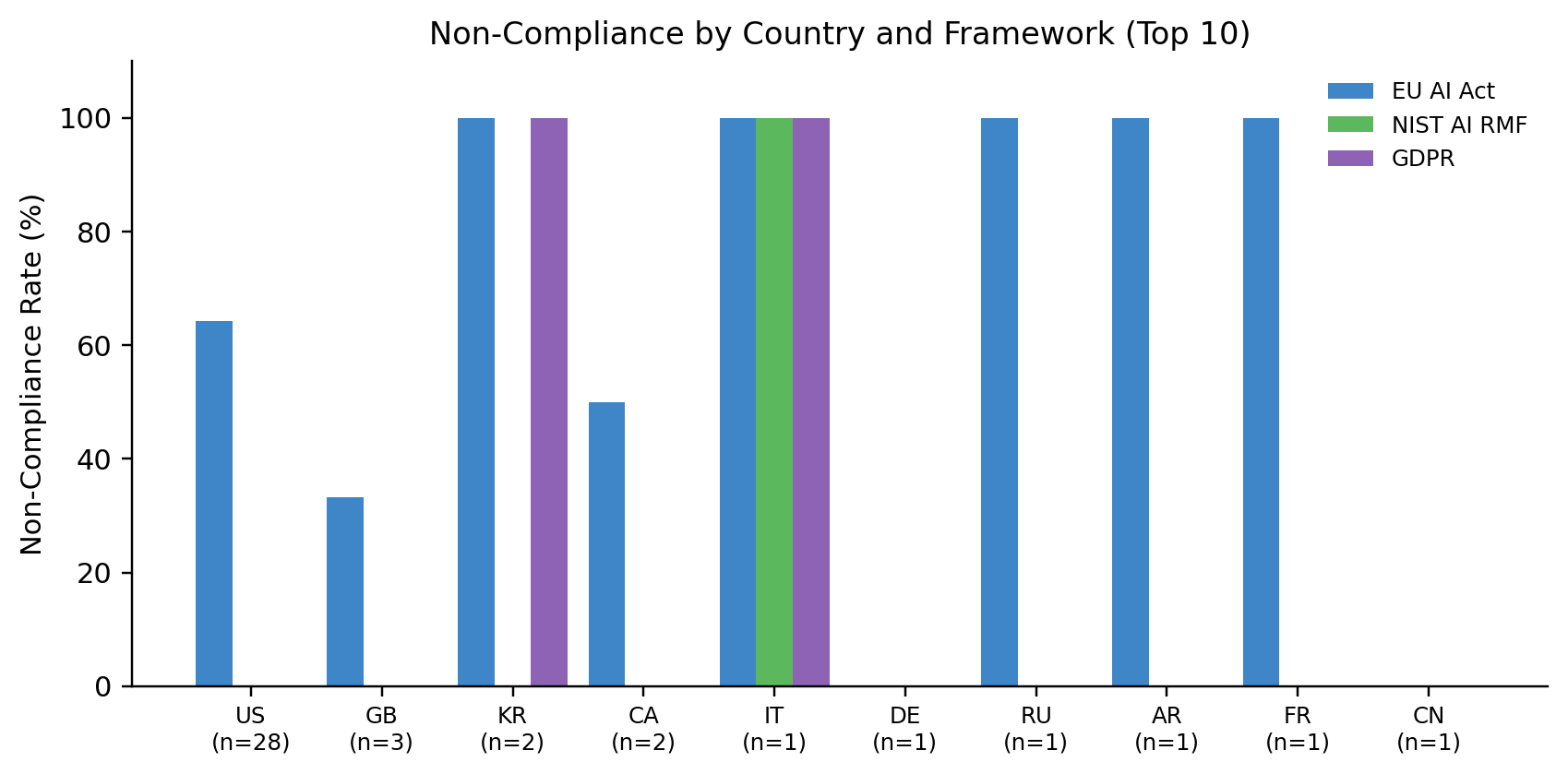}
\caption{Non-compliance by country and framework (top 10 countries).}
\end{figure}

\begin{figure}[t]
\centering
\includegraphics[width=0.9\linewidth]{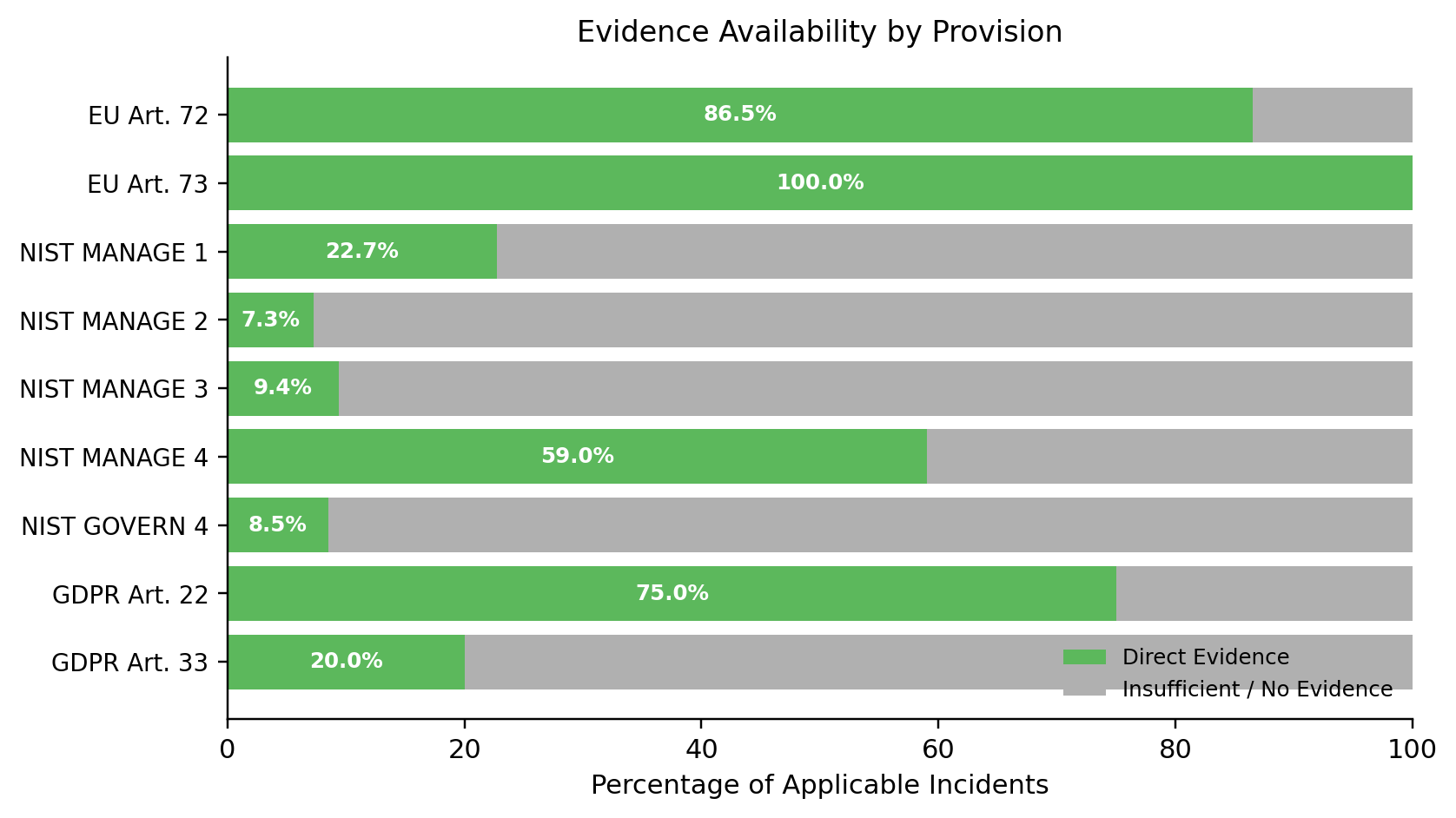}
\caption{Evidence availability by provision.}
\end{figure}

\begin{figure}[t]
\centering
\includegraphics[width=0.9\linewidth]{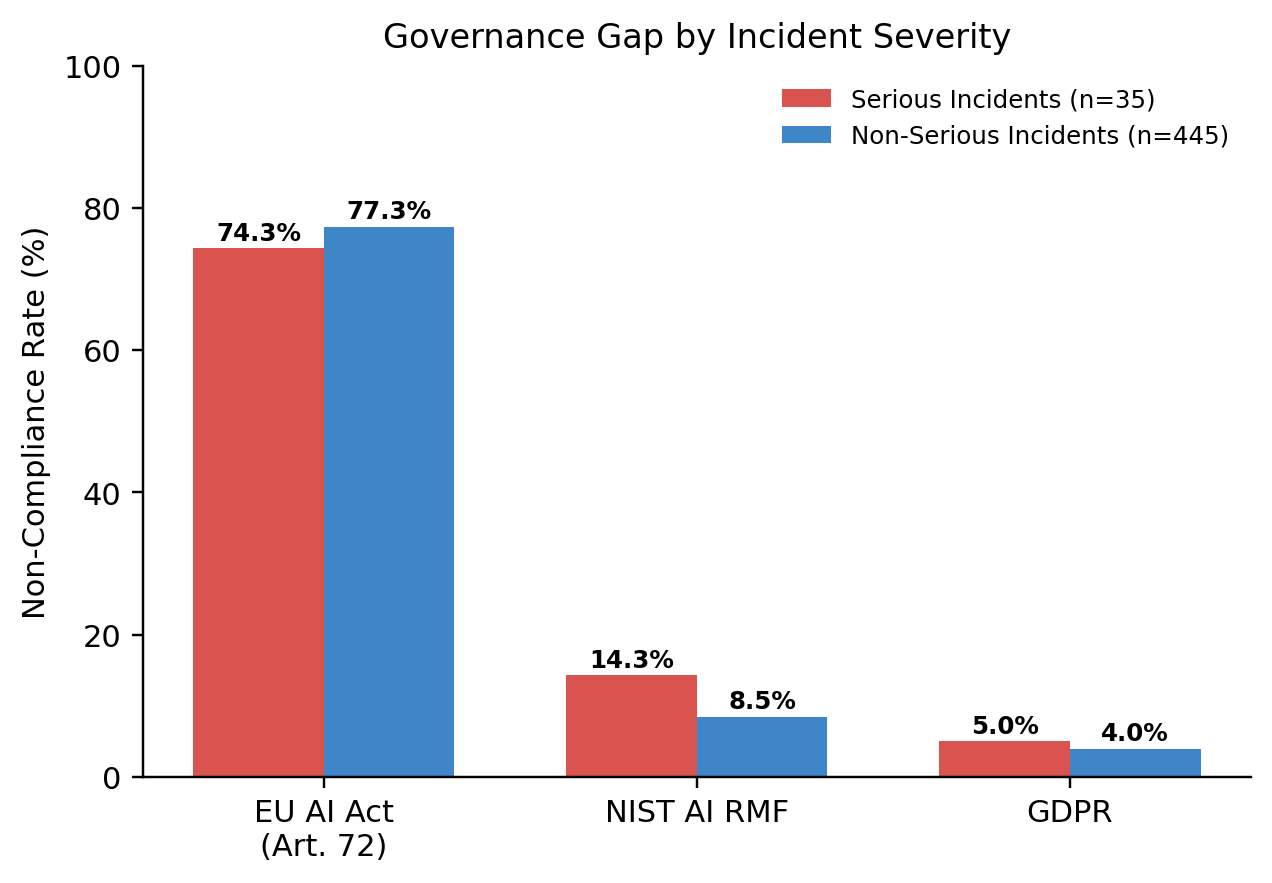}
\caption{Governance gap by incident severity (serious vs non-serious).}
\end{figure}

\begin{figure}[t]
\centering
\includegraphics[width=0.9\linewidth]{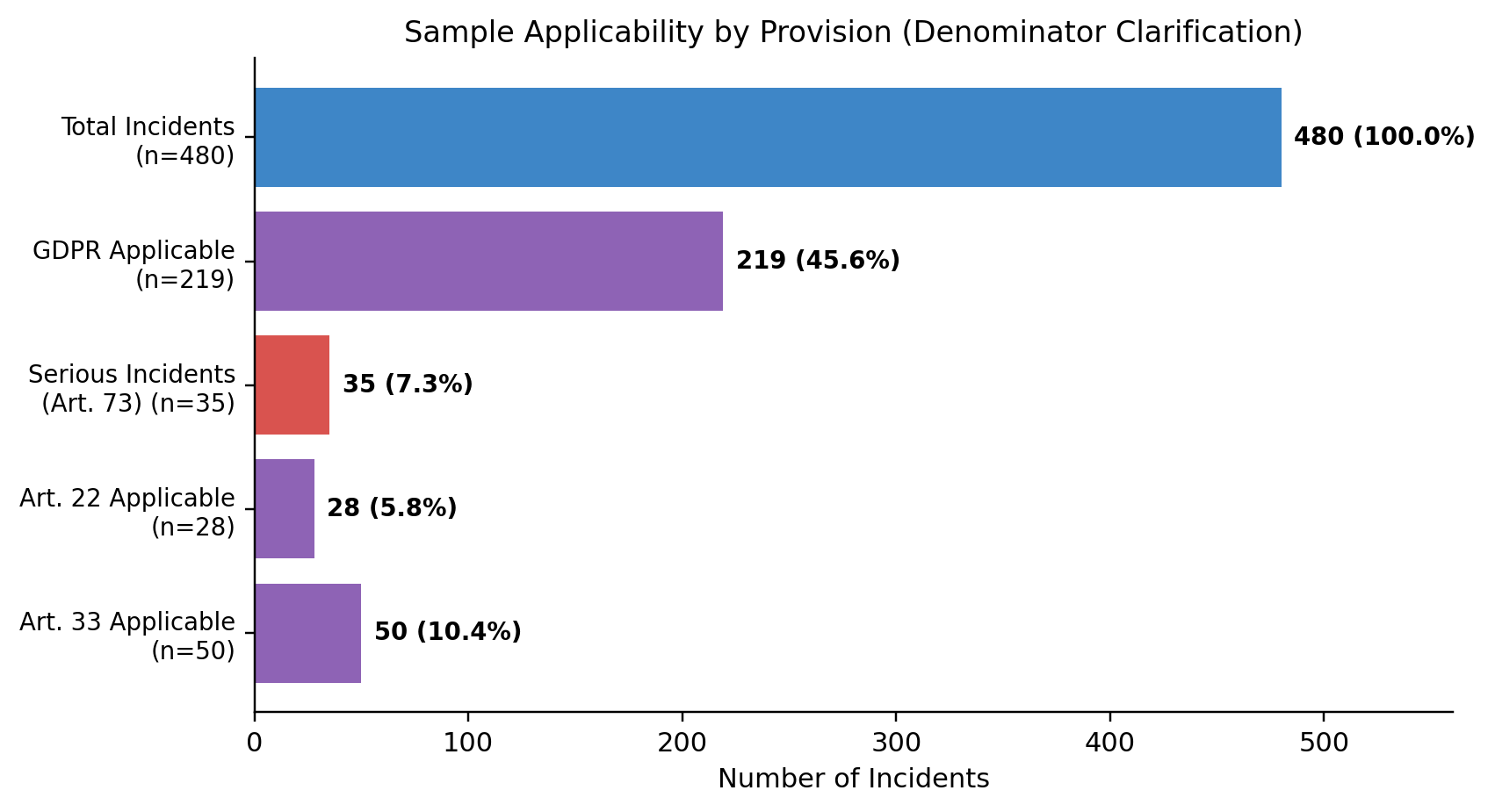}
\caption{Sample applicability by provision (denominator clarification).}
\end{figure}

\section*{Appendix B: PAGCF Implementation Artefacts}
\begin{figure}[t]
\centering
\includegraphics[width=0.9\linewidth]{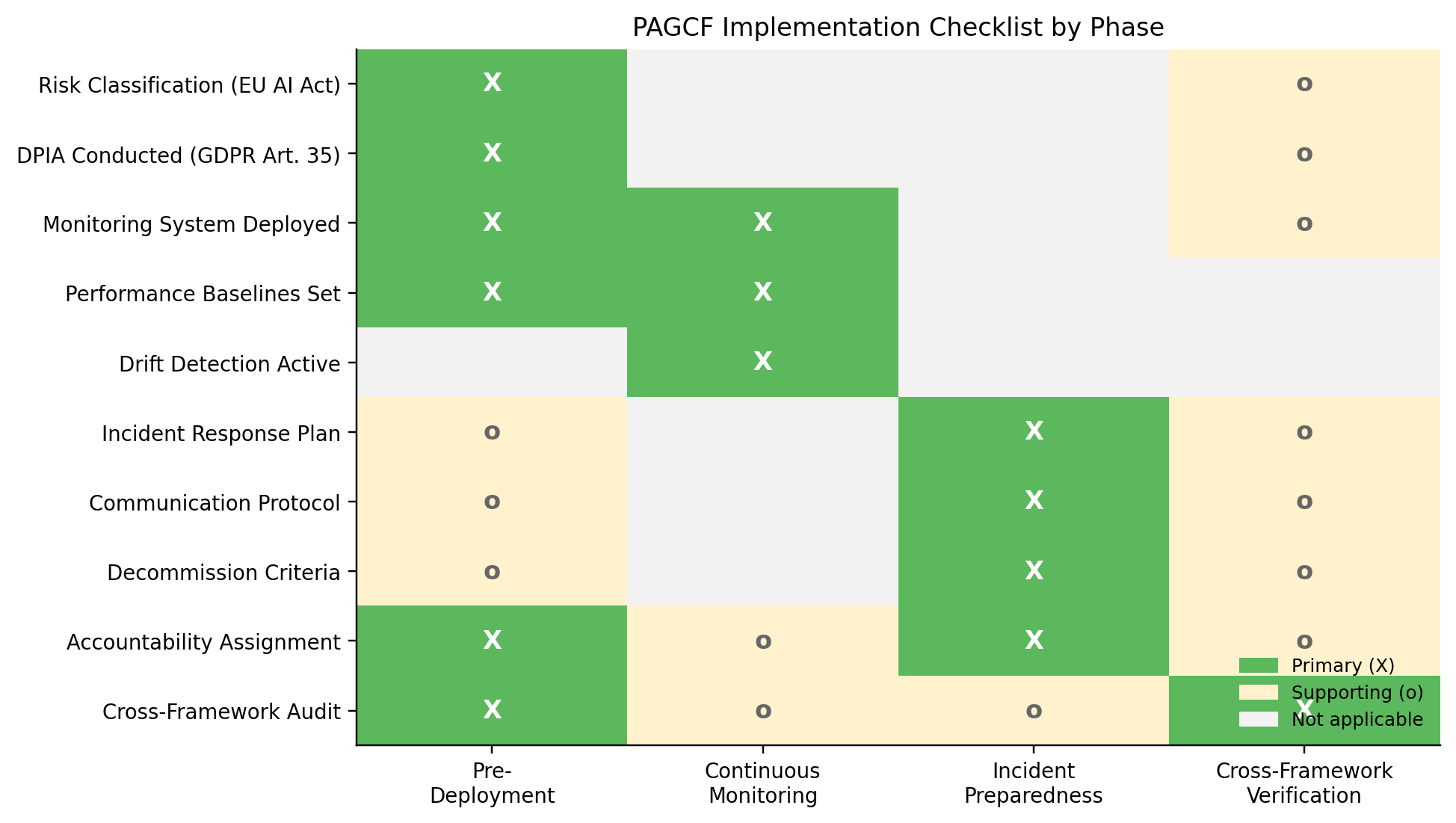}
\caption{PAGCF implementation checklist mapped to framework phases. X = primary; o = supporting.}
\end{figure}

\begin{figure}[t]
\centering
\includegraphics[width=0.9\linewidth]{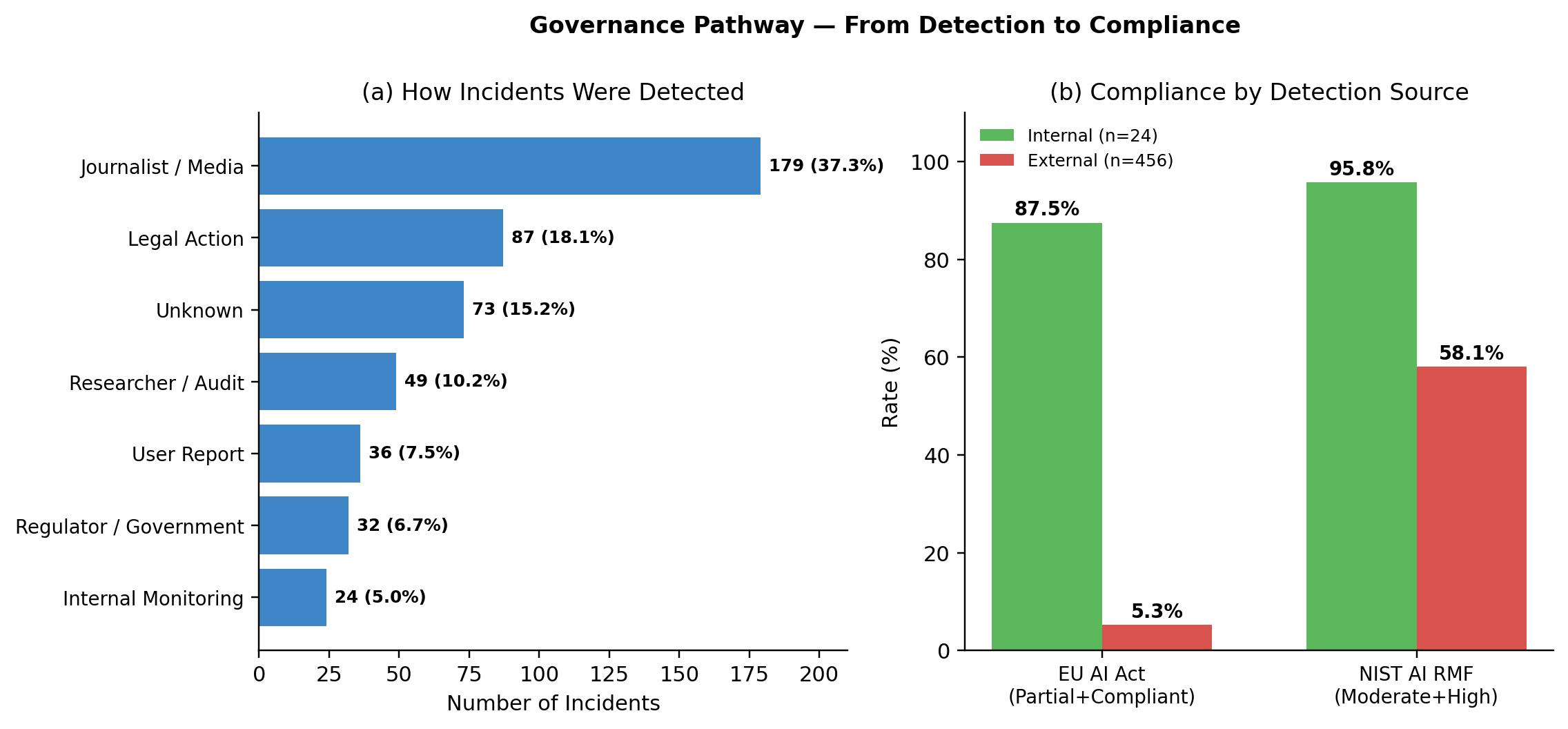}
\caption{Governance pathway from detection to compliance.}
\end{figure}

\end{document}